\documentclass[aps,prl,twocolumn,10pt]{revtex4-2}
\usepackage{amsmath}
\usepackage{graphicx}
\usepackage{hyperref}
\hypersetup{
  colorlinks=true,
  unicode=true,
  linkcolor=blue,
  citecolor=blue,
  urlcolor=blue
}

\begin{document}

\title{Resolving the $\phi$-meson directed-flow puzzle by multi-step meson--baryon dynamics}

\author{
Yingjie Zhou$^{1}$, Taesoo Song$^{1}$, Susanne Gl{\"a}{\ss}el$^{3}$, Jiaxing Zhao$^{7,8}$, Christoph Blume$^{3,1,7}$, Iouri Vassiliev$^{1}$, Vadim Voronyuk$^{4}$, Yaping Wang$^{9}$, Nu Xu$^{9,10}$, J\"org Aichelin$^{5,6}$, Elena Bratkovskaya$^{1,7,8}$
}
\affiliation{$^{1}$ GSI Helmholtzzentrum f\"ur Schwerionenforschung GmbH, Planckstr. 1, 64291 Darmstadt, Germany}
\affiliation{$^{2}$ Institute of Experimental Physics, Heidelberg University, Heidelberg, Germany}
\affiliation{$^{3}$ Institut f\"ur Kernphysik, Max-von-Laue-Str. 1, 60438 Frankfurt, Germany}
\affiliation{$^{4}$ Joint Institute for Nuclear Research, Joliot-Curie 6, 141980 Dubna, Moscow Region, Russia}
\affiliation{$^{5}$ SUBATECH, Nantes University, IMT Atlantique, IN2P3/CNRS, 4 rue Alfred Kastler, 44307 Nantes Cedex 3, France}
\affiliation{$^{6}$ Frankfurt Institute for Advanced Studies, Ruth Moufang Str. 1, 60438 Frankfurt, Germany}
\affiliation{$^{7}$ Helmholtz Research Academy Hessen for FAIR (HFHF), GSI Helmholtz Center for Heavy Ion Physics, Campus Frankfurt, 60438 Frankfurt, Germany}
\affiliation{$^{8}$ Institut f\"ur Theoretische Physik, Johann Wolfgang Goethe University, Max-von-Laue-Str. 1, 60438 Frankfurt, Germany}
\affiliation{$^{9}$ Key Laboratory of Quark and Lepton Physics (MOE) and Institute of Particle Physics, Central China Normal University, Wuhan 430079, China}
\affiliation{$^{10}$ Institute of Modern Physics, Chinese Academy of Sciences, Lanzhou 730000, China}

\begin{abstract} 
Recent STAR measurements at fixed-target Beam Energy Scan energies have
revealed an unexpectedly large directed flow of $\phi$ mesons in Au+Au
collisions, comparable to that of protons and $\Lambda$ baryons and much
stronger than that of light strange mesons. Since the $\phi$ is a
hidden-strangeness meson with relatively weak interactions with
non-strange hadrons, this observation has been interpreted as a possible
signal of unconventional baryonic dynamics or exotic baryonic resonances
coupled to the $\phi$ channel. Within the framework of the Parton-Hadron-Quantum-Molecular-Dynamics(PHQMD) model, we demonstrate that in the high baryon density region, $\phi$ mesons are produced predominantly through multi-step meson--baryon and meson--hyperon
reactions, whose transition amplitudes are constrained by a
coupled-channel $T$-matrix calculation based on an extended SU(6) chiral
effective Lagrangian. Together with the in-medium broadening of the
$\phi$ spectral function, these baryon-driven production channels enhance
near-threshold $\phi$ production and imprint the collective motion of the
baryon-rich source on the produced $\phi$ mesons.
\end{abstract} 

\maketitle


The hidden-strangeness $\phi$ meson is a particularly sensitive probe of the dense stage of relativistic heavy-ion collisions. Owing to its dominantly $s\bar{s}$ structure, long lifetime, and relatively weak interaction with non-strange hadronic matter, the $\phi$ meson is expected to suffer less hadronic rescattering than ordinary light hadrons and may therefore retain information on the early, high-density phase of the reaction. At the same time, in the few-GeV beam-energy domain its production occurs close to threshold, where the system is baryon rich and the microscopic production mechanism becomes essential.

Recently, the STAR Collaboration reported a surprisingly strong directed flow $v_1$ of $\phi$ mesons in Au+Au collisions at $\sqrt{s_{NN}}=3.0$--$4.5$ GeV \cite{STAR:2026ley}.
Directed flow is quantified by the first harmonic coefficient of the
azimuthal distribution with respect to the reaction plane,
$
v_1(y)=\left\langle \cos(\varphi-\Psi_{\rm RP})\right\rangle
      =\left\langle \frac{p_x}{p_T}\right\rangle,
$
and is commonly characterized near midrapidity by the slope
$
F \equiv \left. \frac{dv_1}{dy}\right|_{y=0}.
$
The measured midrapidity slope of the $\phi$-meson directed flow is comparable in
magnitude to that of protons and $\Lambda$ baryons, but is substantially
larger than that of light mesons.

This observation is unexpected in view of the experimentally observed meson--baryon splitting of directed flow at these energies, where baryons and light mesons form two branches with opposite flow patterns. Based on its mesonic identity, the $\phi$ would naively be expected to follow the light-meson branch. Instead, it follows the baryonic branch, exhibiting a directed flow similar to protons and $\Lambda$ baryons. This violation of the simple meson--baryon flow grouping constitutes the $\phi$-meson directed-flow puzzle and points to the importance of the microscopic production mechanism. Earlier STAR measurements at higher Beam Energy Scan energies already demonstrated that strange-hadron directed flow provides stringent constraints on the dynamics of baryon-rich matter~\cite{STAR:2017okv}; the new low-energy data now sharpen this issue for hidden strangeness.

The observed baryon-like $\phi$ flow has motivated interpretations in terms of unconventional production dynamics, including the possible role of high-mass baryonic resonances with sizable couplings to the $\phi$ channel \cite{STAR:2026ley,Steinheimer:2025phiFlow}. In the UrQMD scenario employed for comparison in Ref.~\cite{STAR:2026ley}, the dominant mechanism proceeds through intermediate high-mass nucleon resonances, $N+N \rightarrow N^*+N \rightarrow N+\phi+N $, with states such as $N^*(1990)$, $N^*(2080)$, $N^*(2190)$, $N^*(2220)$, and $N^*(2250)$. In this picture, $\phi$ mesons are produced through baryonic intermediate states and can therefore inherit the collective motion of the baryon-rich medium. The agreement with the STAR data was thus interpreted as evidence that such high-mass $N^*$ resonances play an important role in describing both the enhanced $\phi$ yield and the baryon-like $\phi$-meson directed flow at high baryon density.

In this Letter, we show that the observed baryon-like $\phi$ flow does not require exotic degrees of freedom, hypothetical high-mass resonance decays, or anomalously strong post-production $\phi N$ interactions. Using the microscopic Parton-Hadron-Quantum-Molecular-Dynamics (PHQMD) approach \cite{Aichelin:2019tnk,Coci:2023daq,Kireyeu:2024hjo}, we demonstrate that, at STAR fixed-target Beam Energy Scan energies, $\phi$ mesons are produced predominantly 
through meson--baryon ($mB$) and meson--hyperon ($mY$ where $Y=\Lambda, \Sigma$) reactions rather than through meson--meson scattering. These many-body  multi-step channels, together with the in-medium modification of the $\phi$ spectral function, were previously shown to be essential for describing the enhanced low-energy $\phi$ yield and the large $\phi/K^-$ ratio \cite{Song:2022phi}. The same mechanism naturally explains the directed-flow puzzle: $\phi$ mesons inherit the collective motion of the baryon-rich environment in which they are produced.

Thus, although the $\phi$ is a hidden-strangeness meson, its directed flow at high baryon density is governed primarily by baryon-driven production dynamics. The resulting baryon-like $\phi$-meson $v_1$ is therefore a consequence of conventional hadronic many-body dynamics rather than a direct indication of exotic $\phi$--baryon states.


The Parton-Hadron-Quantum-Molecular Dynamics approach  \cite{Aichelin:2019tnk,Coci:2023daq,Kireyeu:2024hjo} is a microscopic $N$-body transport model that extends the Parton-Hadron-String Dynamics (PHSD) approach \cite{Cassing:2009vt,Moreau:2019vhw} by incorporating the Quantum Molecular Dynamics (QMD) propagation of baryons. In the PHSD--PHQMD framework, the full space-time evolution of relativistic heavy-ion collisions is described, starting from the initial out-of-equilibrium nucleon--nucleon scatterings and string formation, proceeding through the possible creation and evolution of partonic matter, and ending with hadronization and final-state interactions of the produced hadrons. 

In the few-GeV energy regime considered here, the reaction dynamics is dominated by hadronic degrees of freedom and by the collective motion generated in the compressed baryon-rich phase. In PHQMD, baryons are represented by Gaussian wave packets whose mutual interactions are determined by the chosen nuclear equation of state. This provides a genuine $N$-body description of the compressed baryon-rich phase and of the collective dynamics that generates directed flow. In the present study we employ the hard EoS, while soft and soft momentum-dependent options are available for systematic studies.

The production and propagation of $\phi$ mesons are implemented following Ref.~\cite{Song:2022phi}. The baseline production mechanisms include the conventional hadronic reactions commonly used in transport approaches, ordered here as baryon--baryon ($BB$), meson--baryon ($mB$), and meson--meson ($mm$) channels. The key extension introduced in Ref.~\cite{Song:2022phi} is the implementation of additional multi-step meson--baryon and meson--hyperon mechanisms for $\phi$ production. These channels are motivated by coupled-channel $T$-matrix calculations basedon the SU(6) extension of the meson--baryon chiral Lagrangian within a unitary coupled-channel approach~\cite{Gamermann:2011mq}. Such reactions can become important in heavy-ion collisions, where secondary mesons, baryons, and hyperons are abundantly produced, while being absent or strongly suppressed in elementary $p+p$ reactions.

For $\phi$ production from meson--baryon $S$-wave scattering, we include the transition amplitudes in the strangeness-zero sector with
$(I,J)=(1/2,1/2),\ (3/2,1/2),\ (1/2,3/2) ,\ (3/2,3/2), \ (3/2,5/2),$
where $I$ and $J$ denote the total isospin and spin, respectively. The channels considered for $I=1/2$ are
$\eta N, \ K\Lambda, \ K\Sigma, \ \rho N, \ K\Sigma^*, \ \rho\Delta, 
\ K^*\Lambda, \ K^*\Sigma, \ K^*\Sigma^* \rightarrow \phi N ,$
while for the isospin $I=3/2$ they are
$K\Sigma, \ \rho N, \ \eta\Delta, \ K\Sigma^*, \ \rho\Delta, \
K^*\Sigma, \ K^*\Sigma^* \rightarrow \phi\Delta .$
Since the corresponding cross sections are not known experimentally, they are taken from the SU(6)-based coupled-channel $T$-matrix calculation. The inverse reactions are implemented by detailed balance. This provides a microscopic many-body input for the coupling of hidden strangeness to baryonic and hyperonic degrees of freedom in dense hadronic matter.

In-medium effects are included through the spectral functions of vector and pseudoscalar strange mesons also in these additional channels. In particular, the $\phi$ meson acquires an increased collisional width in dense matter due to interactions with the surrounding hadrons. This broadens its spectral function and shifts part of the spectral strength to invariant masses below the vacuum pole mass. 
Near threshold, the available invariant energy in a hadronic collision is only marginally sufficient to produce a $\phi$ meson at its vacuum pole mass, so the production rate is strongly phase-space suppressed. The in-medium broadening of the $\phi$ spectral function shifts part of the spectral strength to lower invariant masses. Such lower-mass $\phi$ excitations require less energy to be produced and therefore open additional phase space, enhancing the effective $\phi$ production probability in the dense medium.

As demonstrated in Ref.~\cite{Song:2022phi}, the many-body motivated
multi-step channels, together with the in-medium modification of the
$\phi$ spectral function, lead to a substantial enhancement of
subthreshold and near-threshold $\phi$ production. This allows one to
describe the measured low-energy $\phi$ yields and the enhanced
$\phi/K^-$ ratio without introducing hypothetical heavy-baryon decays
into the $\phi$ channel. In the present work, this same channel-resolved implementation provides the basis for identifying the microscopic origin of the observed baryon-like $\phi$-meson directed flow.

The propagation of strange mesons in the PHSD--PHQMD framework includes in-medium effects for antikaons and kaons following Ref.~\cite{Song:2020clw}.
For antikaons, the in-medium self-energy is evaluated within a finite-temperature $G$-matrix approach based on a self-consistent unitarization of the transition amplitudes from an effective chiral Lagrangian~\cite{Cabrera:2014lca}. The resulting antikaon spectral function exhibits a collisional broadening and, at low momenta, an attractive mass shift, which lowers the effective production threshold and enhance antikaon production close to threshold. The produced off-shell antikaons are propagated dynamically in the medium. For $K^+$ mesons, a repulsive potential increasing approximately linearly with the nuclear density is employed~\cite{Korpa:2004ae}, which effectively increases the in-medium $K^+$ mass, suppresses its production, and hardens the transverse-momentum spectrum.


\begin{figure}[t!]  
    \centering
    \includegraphics[width=0.99\linewidth]{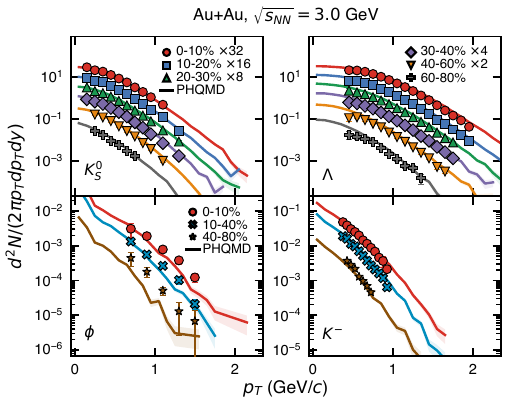}
    \includegraphics[width=0.99\linewidth]{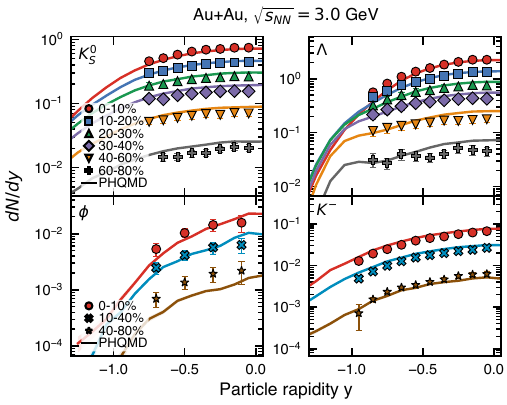}
    \caption{
    Comparison of PHQMD calculations with STAR data
    \cite{STAR:2024znc,STAR:2021hyx} for $K^{0}_{S}$, $\Lambda$,
    $\phi$, and $K^{-}$ production in Au+Au collisions at
    $\sqrt{s_{\rm NN}}=3.0$ GeV. The upper panels show
    transverse-momentum spectra, while the lower panels show rapidity
    density distributions for different centrality classes. Symbols
    denote the data, curves the PHQMD results, and shaded bands the
    statistical uncertainties of the calculations.
    }       
    \label{fig:yield}
\end{figure}
 
For the present analysis, we classify the final $\phi$ mesons according to their production mechanism and group them into meson--meson ($mm$), meson--baryon ($mB$), and baryon--baryon ($BB$) sources, where $mB$ includes both meson--nucleon and meson--hyperon reactions. This channel separation is crucial for identifying the microscopic origin of the $\phi$-meson directed flow. Since the dominant low-energy production mechanism involves a baryon or hyperon in the entrance channel, the produced $\phi$ mesons are expected to reflect the collective motion of the baryon-rich source, even though the $\phi$ itself is a hidden-strangeness meson.

Before addressing the directed-flow puzzle, we first verify that PHQMD provides a good description of the relevant strange-hadron production at $\sqrt{s_{NN}}=3.0$ GeV. As shown in Fig.~\ref{fig:yield}, the calculated transverse-momentum spectra and rapidity distributions of $K^0_S$, $\Lambda$, $\phi$, and $K^-$ are in good agreement with STAR data within the experimental and model uncertainties. This simultaneous description of open and hidden strangeness establishes the production baseline for studying the microscopic origin of the $\phi$-meson directed flow.

\begin{figure}[t!]  
    \centering
    \includegraphics[width=0.99\linewidth]{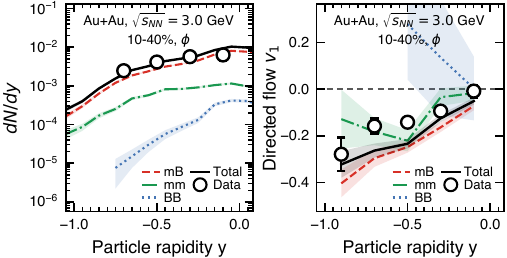}
    \includegraphics[width=0.99\linewidth]{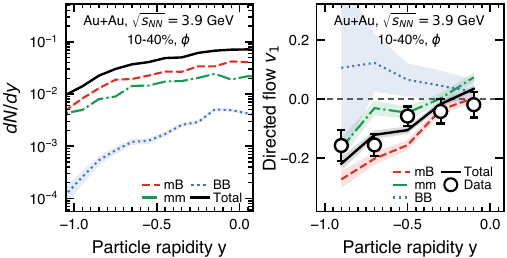}
\caption{
Channel decomposition of the $\phi$-meson rapidity density $dN/dy$ (left) and directed flow $v_1(y)$ (right) in 10--40\% Au+Au collisions at $\sqrt{s_{\rm NN}}=3.0$ GeV (top) and $3.9$ GeV (bottom). The curves show the contributions from baryon--baryon ($BB$), meson--baryon ($mB$, including meson--hyperon ($mY$) reactions), and meson--meson ($mm$) production channels, together with the sum of all channels. Shaded bands denote statistical uncertainties of the calculations. STAR data are shown by symbols~\cite{STAR:2021hyx,STAR:2026ley}.
}
    \label{fig:phigroup}
\end{figure}
\begin{figure*}[!]  
    \centering
    \includegraphics[width=0.9\linewidth]{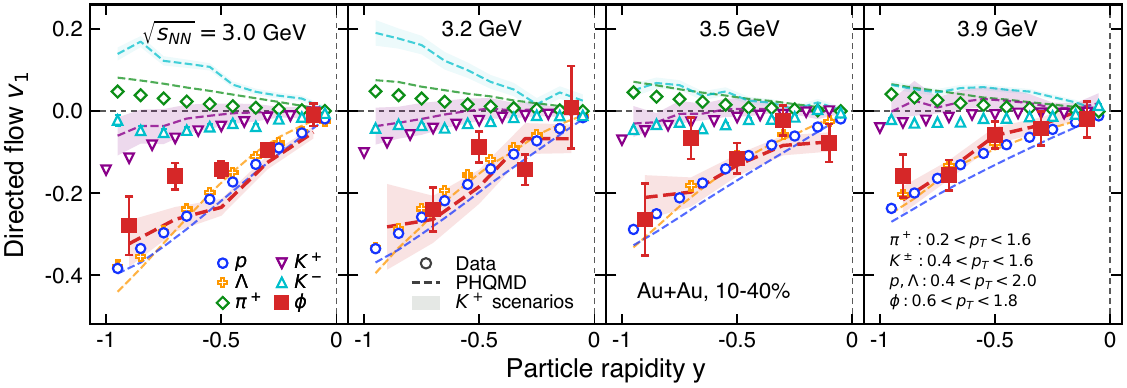}  
\caption{
Directed flow $v_1(y)$ of $\pi^{+}$, $K^{+}$, $K^{-}$, protons,
$\Lambda$ baryons, and $\phi$ mesons in 10--40\% Au+Au collisions at $\sqrt{s_{\rm NN}}=3.0$--$3.9$ GeV. Symbols denote STAR measurements, with statistical and systematic uncertainties shown by bars and boxes, respectively~\cite{STAR:2026ley,STAR:2025twg}. Dashed curves show the PHQMD calculations. The shaded bands for the $\phi$ meson represent the statistical uncertainties of the model calculations. The shaded bands for $K^+$ indicate the model uncertainty associated with the in-medium kaon potentials, corresponding to a variation of the repulsive $K^+$ potential between $U_{K^+}=0$ and $U_{K^+}\simeq25$ MeV at normal nuclear density, including the statistical uncertainties of both calculations.}
\label{fig:v1}
\end{figure*}

The microscopic origin of the baryon-like $\phi$ flow is exposed by classifying the final $\phi$ mesons according to their production channel. Figure~\ref{fig:phigroup} shows the channel decomposition of the rapidity density and directed flow at $\sqrt{s_{NN}}=3.0$ and $3.9$ GeV. At both energies, the $\phi$ yield is dominated by meson--baryon reactions, including meson--hyperon channels, while the meson--meson contribution remains subleading. The same channel hierarchy is reflected in the directed flow: the $mB$ component carries a strong baryon-like $v_1$ and largely determines the total $\phi$-meson flow. Thus, the observed similarity between the $\phi$ and baryon directed flows does not require the $\phi$ meson to undergo strong baryon-like rescattering after its production. Instead, it follows from the baryon-driven environment in which most $\phi$ mesons are produced.

The central result of this work is presented in Fig.~\ref{fig:v1}, where
PHQMD calculations are compared with recent STAR measurements of
$v_1(y)$ for $\pi^+$, $K^+$, $K^-$, protons, $\Lambda$ baryons, and
$\phi$ mesons in 10--40\% Au+Au collisions at
$\sqrt{s_{\rm NN}}=3.0$--$3.9$ GeV. PHQMD provides a good overall
description of the measured directed flow for the different hadron
species, although it somewhat overestimates the magnitude of the $K^-$
flow. Most importantly, the calculation reproduces the large absolute
magnitude of the $\phi$-meson $v_1$. Baryons, represented by protons and
$\Lambda$'s, exhibit a strong directed flow, whereas pions and charged
kaons show substantially smaller absolute values of $v_1$. The $\phi$
meson, however, follows the baryonic branch rather than the light-meson
one.

This difference reflects the distinct microscopic production mechanisms
of charged kaons and $\phi$ mesons. The charged-kaon flow is shaped by a
mixture of direct meson--baryon production, rescattering, in-medium
effects, and $K^*$ decays, which partially wash out the initial flow
anisotropy and lead to a small final magnitude of $v_1$. This is
especially relevant for $K^+$ mesons, where a sizable contribution from
$K^*$ decays is already present at $\sqrt{s_{\rm NN}}=3.0$ GeV and
becomes dominant at $3.9$ GeV. In contrast, most $\phi$ mesons are
produced through meson--baryon and meson--hyperon channels that retain
the collective motion of the baryon-rich source. Combined with the
channel decomposition in Fig.~\ref{fig:phigroup}, Fig.~\ref{fig:v1}
demonstrates that the baryon-like $\phi$ flow originates from its
production dynamics rather than from its identity as a hidden-strangeness
meson.


In summary, we have demonstrated that the strong directed flow of $\phi$ mesons recently observed at high baryon density can be understood within microscopic hadronic many-body dynamics. PHQMD provides a good description of the measured $\phi$-meson collective flow $v_1$ at STAR fixed-target Beam Energy Scan energies and identifies its microscopic origin through a channel-resolved analysis. The key point is that, at these energies, $\phi$ mesons are produced predominantly in multi-step meson--baryon and meson--hyperon reactions. Consequently, the produced $\phi$ mesons carry the collective motion of the baryon-rich environment in which they are generated, leading naturally to a baryon-like directed flow. In addition, the same mechanism also reproduce the yield ratio of $\phi/K^{-}$.

This mechanism may resolve the apparent $\phi$-meson directed-flow puzzle without invoking exotic baryonic resonances, hypothetical heavy-resonance decays into the $\phi$ channel, or anomalously strong post-production $\phi N$ interactions. Although the $\phi$ is a hidden-strangeness meson, its flow at these energies is governed primarily by the baryonic origin of its production rather than by the flow pattern of light mesons. The $\phi$ meson thus provides not only a probe of hidden strangeness, but also a direct diagnostic of the microscopic production dynamics in dense baryonic matter.

\begin{acknowledgments}
We thank M. Bleicher and J. Steinheimer for valuable discussions. This work is supported by the FAIR Fellowship and Associate Program of GSI Helmholtzzentrum für Schwerionenforschung, Darmstadt, Germany, the National Natural Science Foundation of China under Grant No. 12375134, the National Key Research and Development Program of China (Grant No. 2024YFE0110103 and 2024YFA1611003), and the Fundamental Research Funds for the Central Universities (Grant No. CCNU25JCPT017).
\end{acknowledgments}

\bibliography{phqmdref} 

\end{document}